**The Past, Present, and Future of the Brain Imaging Data Structure (BIDS)**

Poldrack, Russell A.[1], Markiewicz, Christopher J.[1], Appelhoff, Stefan[2], Ashar, Yoni K.[3], Auer, Tibor[4,5], Baillet, Sylvain[6], Bansal, Shashank[7], Beltrachini, Leandro[8], Benar, Christian G.[9], Bertazzoli, Giacomo[10,11,12,13,14], Bhogawar, Suyash[15], Blair, Ross W.[1], Bortoletto, Marta[10], Boudreau, Mathieu[16], Brooks, Teon L.[1], Calhoun, Vince D.[17], Castelli, Filippo Maria[18,19], Clement, Patricia[20,21], Cohen, Alexander L.[22], Cohen-Adad, Julien[16], D'Ambrosio, Sasha[23,24], de Hollander, Gilles[25], de la Iglesia-Vayá, María[26], de la Vega, Alejandro[27], Delorme, Arnaud[28], Devinsky, Orrin[29], Draschkow, Dejan[30], Duff, Eugene Paul[31], DuPre, Elizabeth[1], Earl, Eric[32], Esteban, Oscar[33], Feingold, Franklin W.[1], Flandin, Guillaume[34], Galassi, Anthony[32], Gallitto, Giuseppe[35,36], Ganz, Melanie[37,38], Gau, Rémi[39], Gholam, James[40], Ghosh, Satrajit S.[41], Giacomel, Alessio[42], Gillman, Ashley G.[43], Gleeson, Padraig[44], Gramfort, Alexandre[45], Guay, Samuel[46], Guidali, Giacomo[47], Halchenko, Yaroslav O.[48], Handwerker, Daniel A.[32], Hardcastle, Nell[1], Herholz, Peer[49], Hermes, Dora[50], Honey, Christopher J.[51], Innis, Robert B.[32], Ioanas, Horea-Ioan[48], Jahn, Andrew[52], Karakuzu, Agah[16], Keator, David B.[53,54,55], Kiar, Gregory[56], Kincses, Balint[35,36], Laird, Angela R.[57], Lau, Jonathan C.[58], Lazari, Alberto[59], Legarreta, Jon Haitz[60], Li, Adam[61], Li, Xiangrui[62], Love, Bradley C.[63], Lu, Hanzhang[64], Marcantoni, Eleonora[65], Maumet, Camille[66], Mazzamuto, Giacomo[67], Meisler, Steven L.[68], Mikkelsen, Mark[69], Mutsaerts, Henk[70,71], Nichols, Thomas E.[72], Nikolaidis, Aki[73], Nilsonne, Gustav[74,75], Niso, Guiomar[76], Norgaard, Martin[32,37], Okell, Thomas W.[59], Oostenveld, Robert[77,78], Ort, Eduard[79], Park, Patrick J.[80], Pawlik, Mateusz[81], Pernet, Cyril R.[38], Pestilli, Franco[27], Petr, Jan[82], Phillips, Christophe[83], Poline, Jean-Baptiste[84], Pollonini, Luca[85,86], Raamana, Pradeep Reddy[87], Ritter, Petra[88,89,90,91,92], Rizzo, Gaia[93,94], Robbins, Kay A.[95], Rockhill, Alexander P.[96], Rogers, Christine[97], Rokem, Ariel[98], Rorden, Chris[99], Routier, Alexandre[100], Saborit-Torres, Jose Manuel[26], Salo, Taylor[101], Schirner, Michael[88,89,90,91,92], Smith, Robert E.[102,103], Spisak, Tamas[35,104], Sprenger, Julia[105], Swann, Nicole C.[106], Szinte, Martin[105], Takerkart, Sylvain[105], Thirion, Bertrand[45], Thomas, Adam G.[32], Torabian, Sajjad[107], Varoquaux, Gael[108], Voytek, Bradley[109], Welzel, Julius[110], Wilson, Martin[111], Yarkoni, Tal[112], Gorgolewski, Krzysztof J.[1]

1: Department of Psychology, Stanford University, Stanford, CA, USA
2: Max Planck Institute for Human Development, Berlin, Germany
3: University of Colorado Anschutz Medical Campus, Aurora, CO, USA
4: School of Psychology, University of Surrey, Guildford, UK
5: Artificial Intelligence and Informatics group, Rosalind Franklin Institute, Harwell Campus, Didcot, UK
6: McConnell Brain Imaging Centre, Montréal Neurological Institute, McGill University, Montréal, Canada
7: Department of Bioengineering, University of California, San Diego, La Jolla, CA, USA
8: Cardiff University Brain Research Imaging Centre (CUBRIC), School of Physics and Astronomy, Cardiff University, Wales, UK
9: Aix Marseille Université, INSERM, INS, Inst Neurosci Syst, Marseille, France
10: Neurophysiology Lab, IRCCS Istituto Centro San Giovanni di Dio Fatebenefratelli, Brescia, Italy
11: Center for Mind/Brain Sciences - CIMeC, University of Trento, Rovereto, TN, Italy

12: Brigham and Women’s Hospital, Boston, MA, USA
13: Massachusetts General Hospital, Boston, MA, USA
14: Harvard Medical School, Boston, MA, USA
15: Rackspace Technology, San Antonio, TX, USA
16: NeuroPoly Lab, Polytechnique Montréal, Montréal, Quebec, Canada
17: Tri-institutional Center for Translational Research in Neuroimaging and Data Science (TReNDS), Georgia State, Georgia Tech, Emory, Atlanta, GA, USA
18: European Laboratory for Non-Linear Spectroscopy (LENS), University of Florence, Sesto Fiorentino, Italy
19: Bioretics srl, Cesena, Italy
20: Department of Medical Imaging, Ghent University Hospital, Ghent, Belgium
21: Department of Diagnostic Sciences, Ghent University, Ghent, Belgium
22: Department of Neurology, Boston Children's Hospital, Boston, MA, USA
23: Dipartimento di Scienze della Salute dell'Università degli Studi di Milano, Milan, Italy
24: Department of Clinical and Experimental Epilepsy, University College London, UK
25: Zurich Center for Neuroeconomics, Department of Economics, University of Zurich, Zurich, Switzerland
26: UMIB-FISABIO, Valencia, Spain
27: The University of Texas at Austin, Austin, TX, USA
28: SCCN, University of California, San Diego, La Jolla CA USA
29: Department of Neurology, NYU Langone Medical Center, New York, NY, USA
30: Department of Experimental Psychology, University of Oxford, Oxford, UK
31: UK Dementia Research Institute, Department of Brain Sciences, Imperial College London, London, UK
32: Intramural Research Program, National Institute of Mental Health, Bethesda, MD, USA
33: Department of Radiology, Lausanne University Hospital and University of Lausanne, Lausanne, Switzerland
34: Wellcome Centre for Human Neuroimaging, University College London, London, England, UK
35: Center for Translational Neuro- and Behavioral Sciences, University Medicine Essen, Essen, Germany
36: Department of Neurology, University Medicine Essen, Essen, Germany
37: Department of Computer Science, University of Copenhagen, Copenhagen, Denmark
38: Neurobiology Research Unit, Copenhagen University Hospital, Copenhagen, Denmark
39: Origamin Lab, The Neuro, McGill University, Montreal, Quebec, Canada
40: Cardiff University Brain Research Imaging Centre (CUBRIC), School of Psychology, Cardiff University, Wales, UK
41: Massachusetts Institute of Technology, Cambridge, MA, USA
42: Department of Neuroimaging, Institute of Psychiatry, Psychology and Neuroscience, King’s College London, London, England, UK
43: The Australian e-Health Research Centre, Commonwealth Scientific and Industrial Research Organisation, Townsville, Queensland, Australia
44: Department of Neuroscience, Physiology and Pharmacology, University College London, London, England, UK

45: Inria, CEA, Université Paris-Saclay, Palaiseau, France
46: Université de Montréal, Montréal, QC, Canada
47: Department of Psychology & NeuroMI - Milan Centre for Neuroscience, University of Milano-Bicocca, Milan, Italy
48: Center for Open Neuroscience, Department of Psychological and Brain Sciences, Dartmouth College, NH, USA
49: McConnell Brain Imaging Centre, Montréal Neurological Institute, McGill University, Montréal, Quebec, Canada
50: Department of Physiology and Biomedical Engineering, Mayo Clinic, Rochester, MN, USA
51: Department of Psychological & Brain Sciences, Johns Hopkins University, Baltimore, MD, USA
52: Functional MRI Laboratory, University of Michigan, Ann Arbor, MI, USA
53: Change Your Brain Change Your Life Foundation, Costa Mesa, CA, USA
54: Amen Clinics, Costa Mesa, CA, USA
55: Department of Psychiatry and Human Behavior, School of Medicine, University of California, Irvine, CA, USA
56: Center for Data Analytics, Innovation, and Rigor, Child Mind Institute, New York, NY USA
57: Department of Physics, Florida International University, Miami, FL, USA
58: Department of Clinical Neurological Sciences, Western University, London, Ontario, Canada
59: Wellcome Centre for Integrative Neuroimaging, FMRIB, Nuffield Department of Clinical Neurosciences, University of Oxford, Oxford, UK
60: Department of Radiology, Brigham and Women's Hospital, Mass General Brigham/Harvard Medical School, Boston, MA, USA
61: Columbia University, New York, NY, USA
62: Center for Cognitive and Behavioral Brain Imaging, The Ohio State University, Columbus, OH, USA
63: University College London, London, UK
64: Johns Hopkins University School of Medicine, Baltimore, MD, USA
65: School for Psychology and Neuroscience and Centre for Cognitive Neuroimaging, University of Glasgow, Glasgow
66: Inria, Univ Rennes, CNRS, Inserm, IRISA UMR 6074, Empenn ERL U 1228, Rennes, France
67: National Research Council - National Institute of Optics (CNR-INO), Florence, Italy
68: Program in Speech and Hearing Bioscience and Technology, Harvard University, Cambridge, MA, USA
69: Department of Radiology, Weill Cornell Medicine, New York, NY, USA
70: Radiology and Nuclear Medicine, Vrije Universiteit Amsterdam, Amsterdam UMC location VUmc, Amsterdam, The Netherlands
71: Amsterdam Neuroscience, Brain Imaging, Amsterdam, The Netherlands
72: Big Data Institute, Li Ka Shing Centre for Health Information and Discovery, Nuffield Department of Population Health, University of Oxford, Oxford, UK
73: Center for the Developing Brain, Child Mind Institute, New York, NY, USA
74: Department of Clinical Neuroscience, Karolinska Institutet, Stockholm, Sweden
75: Swedish National Data Service, Gothenburg University, Gothenburg, Sweden

76: Instituto Cajal, CSIC, Madrid, Spain
77: Donders Institute for Brain, Cognition and Behaviour, Radboud University Nijmegen, Nijmegen, The Netherlands
78: NatMEG, Karolinska Institutet, Stockholm, Sweden
79: Heinrich Heine University, Department of Biological Psychology of Decision Making, Düsseldorf, Germany
80: Western University, London, Ontario, Canada
81: Paris-Lodron-University of Salzburg, Department of Psychology, Centre for Cognitive Neuroscience, Salzburg, Austria
82: Helmholtz-Zentrum Dresden-Rossendorf, Institute of Radiopharmaceutical Cancer Research, Dresden, Germany
83: GIGA CRC in vivo imaging, Liege University, Liege, Belgium
84: Neuro Data Science ORIGAMI Laboratory, McConnell Brain Imaging Centre, Faculty of Medicine, McGill University, Montréal, Canada
85: Department of Engineering Technology, University of Houston, Houston, TX
86: Basque Center on Cognition, Brain and Language, Donostia-San Sebastián, Spain
87: University of Pittsburgh, Pittsburgh, PA, USA
88: Berlin Institute of Health at Charité, Universitätsmedizin Berlin, Charitéplatz 1, Berlin 10117, Germany
89: Department of Neurology with Experimental Neurology, Charité, Universitätsmedizin Berlin, Corporate member of Freie Universität Berlin and Humboldt Universität zu Berlin, Charitéplatz 1, Berlin 10117, Germany
90: Bernstein Focus State Dependencies of Learning and Bernstein Center for Computational Neuroscience, Berlin, Germany
91: Einstein Center for Neuroscience Berlin, Charitéplatz 1, Berlin 10117, Germany
92: Einstein Center Digital Future, Wilhelmstraße 67, Berlin 10117, Germany
93: Invicro, London, UK
94: Division of Brain Sciences, Imperial College London, London, UK
95: Department of Computer Science, University of Texas at San Antonio, San Antonio, TX, USA
96: Department of Neurosurgery, Oregon Health & Science University, Portland, OR, USA
97: McGill Centre for Integrative Neuroscience (MCIN), Montréal Neurological Institute, McGill University, Montréal, QC, Canada
98: University of Washington, Department of Psychology and eScience Institute, Seattle, WA, USA
99: University of South Carolina, Department of Psychology, Columbia, SC, USA
100: Inria, Institut du Cerveau - Paris Brain Institute, Paris, France
101: Lifespan Informatics and Neuroimaging Center (PennLINC), Department of Psychiatry, Perelman School of Medicine, University of Pennsylvania, Philadelphia, PA, USA
102: The Florey Institute of Neuroscience and Mental Health, Heidelberg, Victoria, Australia
103: The Florey Department of Neuroscience and Mental Heath, The University of Melbourne, Parkville, Victoria, Australia
104: Institute for Diagnostic and Interventional Radiology and Neuroradiology, University Medicine Essen, Essen, Germany

105: Institut de Neurosciences de la Timone (INT), UMR7289, CNRS, Aix-Marseille Université, France
106: University of Oregon, Department of Human Physiology, Eugene, OR, USA
107: University of California, Irvine, CA, USA
108: Inria, Université Paris Saclay, Saclay, France
109: Department of Cognitive Science, Halıcıoğlu Data Science Institute, and Neurosciences Graduate Program, University of California, San Diego, La Jolla, CA, USA
110: Kiel University, Kiel, Germany
111: University of Birmingham, Centre for Human Brain Health and School of Psychology, Birmingham, UK
112: Google X. Mountain View, CA USA

## Abstract

The Brain Imaging Data Structure (BIDS) is a community-driven standard for the organization of data and metadata from a growing range of neuroscience modalities. This paper is meant as a history of how the standard has developed and grown over time. We outline the principles behind the project, the mechanisms by which it has been extended, and some of the challenges being addressed as it evolves. We also discuss the lessons learned through the project, with the aim of enabling researchers in other domains to learn from the success of BIDS.

## Main text

The sharing of scientific research data is beneficial in numerous ways. Foremost, it maximizes the potential knowledge to be derived from the data, thus maximizing the benefit to the stakeholders who fund the research and the contributions of research participants. It also provides the means for researchers to attempt to reproduce the work of others in their field, which is an essential component of science. Further, it levels the scientific playing field by providing data to researchers from under-resourced environments or those without data acquisition capabilities, which they can use to develop novel analysis methods or test new scientific hypotheses. Given the rise of machine learning methods in science, another unsung benefit of data sharing is that it provides larger and more diverse training datasets, which can increase robustness and decrease overfitting and bias. For all of these reasons, the sharing of data has become increasingly common across science, as have demands from funding agencies and publishers that data be shared. Within the field of neuroimaging, data sharing efforts started around 2000 with the fMRI Data Center (Van Horn et al., 2013). Ten years later it began to flourish with the advent of the International Neuroimaging Data-sharing Initiative/Functional Connectomes Project (INDI/FCP) (Mennes et al., 2013). Subsequent projects have focused on prospective sharing of data, including the Human Connectome Project (Van Essen et al., 2013) and the Adolescent Brain Cognitive Development (ABCD) Study (Casey et al., 2018), which have had a major impact on the field by providing large quantities of neuroimaging data to researchers.

The sharing of data is a worthy goal, but only if the data are shared in a way that makes them FAIR (Findable, Accessible, Interoperable, and Reusable) (Wilkinson et al., 2016). One major contributor to FAIRness is the use of standard file formats, which allow researchers to reuse data across multiple software platforms. The field of neuroimaging research using MRI has benefitted from the longstanding convergence of the field on a standard file format for imaging data, the Neuroimaging Informatics Technology Initiative (NIfTI) format (Cox, R. W., Ashburner, J., Breman, H., & Fissell, K., 2004), which is used by nearly all major MRI analysis software packages. However, beyond file formats, the data and associated metadata must be organized in such a way that data recipients can quickly and accurately understand the contents of the data. The benefits of a clear, formal organization scheme for data include minimizing the burden of curation for researchers and for data sharing repositories, reducing the likelihood of errors due to misunderstanding or misinterpretation of the data, enabling the development of analysis tools that can automatically utilize the structure of the data to analyze them appropriately and with minimal user input, and affording the ability to automatically validate the data to determine whether they meet the standard (Gorgolewski et al., 2016).

In this paper we outline the history, current status, and future directions of the Brain Imaging Data Structure (BIDS), which has become a widely accepted community-driven data standard within the neuroscience community. The goal of this exposition is to provide a written history of the events that gave rise to BIDS and outline the events leading to its establishment, which we hope will be of interest to researchers in the neuroimaging field as well as to those in other fields working to establish successful new data standards.

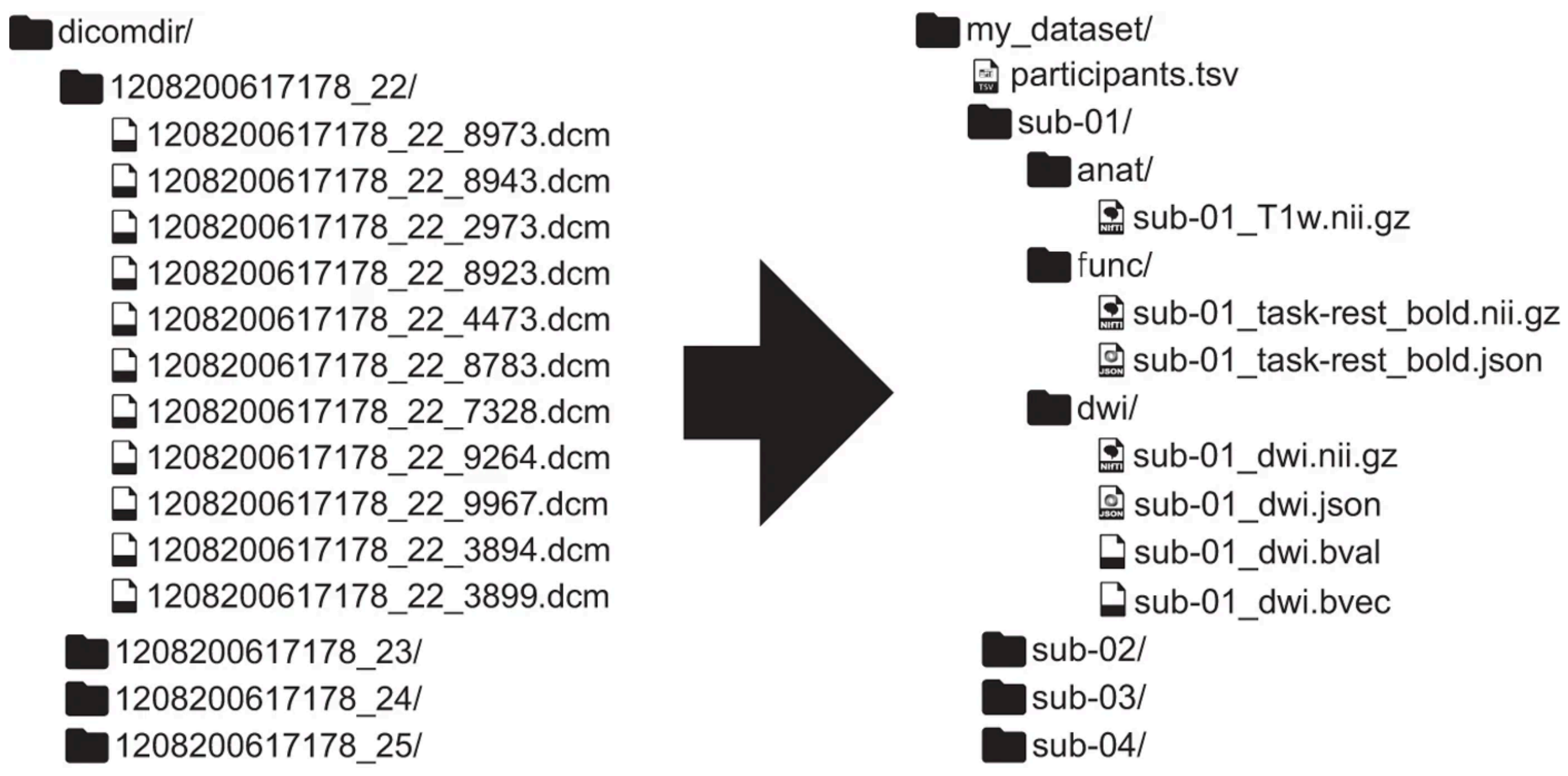

**Figure 1.** A reproduction of Gorgolewski et al., 2016, Figure 1, showing an example mapping from DICOM to BIDS. BIDS is a community-driven standard for organizing, naming, and annotating neuroimaging data that places a heavy emphasis on human- and machine-readability. Since its initial publication, BIDS has expanded from structural, functional

and diffusion MRI to incorporate other MR methodologies such as arterial spin labeling and other recording technologies such as electrophysiology.

### The birth of BIDS

The birth of BIDS can be traced back ultimately to a social media post by Russ Poldrack on October 17, 2014[1]. The post referred to a talk that Chris Gorgolewski had given at the weekly Stanford cognitive/neuroscience seminar ("Frisem"). A reply to the post by Stuart Buck, then a program officer at the Laura and John Arnold Foundation, led to a discussion that ultimately resulted in a substantial grant from the Foundation to the Stanford group, with the aim of developing a new data sharing platform that would supercede the OpenfMRI data sharing platform (Poldrack et al., 2013) that the group had previously run. This new platform would ultimately become the OpenNeuro archive (Markiewicz et al., 2021) and the support from the Arnold Foundation would help start the work on BIDS.

One of the major challenges of the OpenfMRI project had been data curation. The project had developed an in-house data organization scheme (see Figure 1 in Poldrack et al., 2013), which reflected common practice in many labs but was built around a specific workflow for task fMRI analysis based on the FSL software package (Jenkinson et al., 2012). Standardization of file layout and study design metadata gained traction in other projects interested in formalizing the loading of data; e.g., PyMVPA (Hanke et al., 2009) version 2.6.1 (November 2014), included "Direct support for loading data and design from openfmri.org-style datasets"[2]. Since the scheme was not formally described, researchers wishing to deposit data could not easily transform their data to meet it. Instead, data depositors would send their data to the OpenfMRI team and a curator within the team would work with the depositor to transform the data to meet the informal standard. There was also no way to easily determine whether this transformation was correct, other than running it through the automated workflow and seeing whether the workflow ran successfully. This resulted in significant personnel costs and greatly limited the amount of data that could be ingested to the archive. This standard was also limited to a very specific type of MRI data and analysis workflow, and thus was not necessarily useful for a broad group of researchers.

When Chris Gorgolewski and Russ Poldrack began discussing the development of a new archive, they recognized that it was essential to substantially shift the burden of curation from the archive to the data owners, in order to make the archive financially sustainable in the long run. It became clear that this would require a detailed and general scheme for the organization of the intended data types, and that this scheme should support automated validation so that users can upload data and share them immediately without the need for manual curation. As it happened, the Stanford team was already participating in the Neuroimaging Data Sharing Task Force (or NIDASH), a collective effort under the umbrella of the International Neuroinformatics Coordinating Facility (INCF). Amongst other initiatives, this group was developing the Neuroimaging Data Model (NIDM) project, an informatics framework for the formal description of neuroimaging experiments and datasets (Maumet et al., 2016). As part of a series of meetings organized by INCF to further the progress of NIDM and other data sharing efforts, a meeting had been planned for January 2015 at Stanford in order to discuss the development of a NIDM model for fMRI experiments, based on the OpenfMRI use case. This was the meeting at which BIDS was first envisioned.

---

[1] https://twitter.com/russpoldrack/status/523263185764491264; screenshot available at https://osf.io/ha8gx

[2] https://github.com/PyMVPA/PyMVPA/pull/240

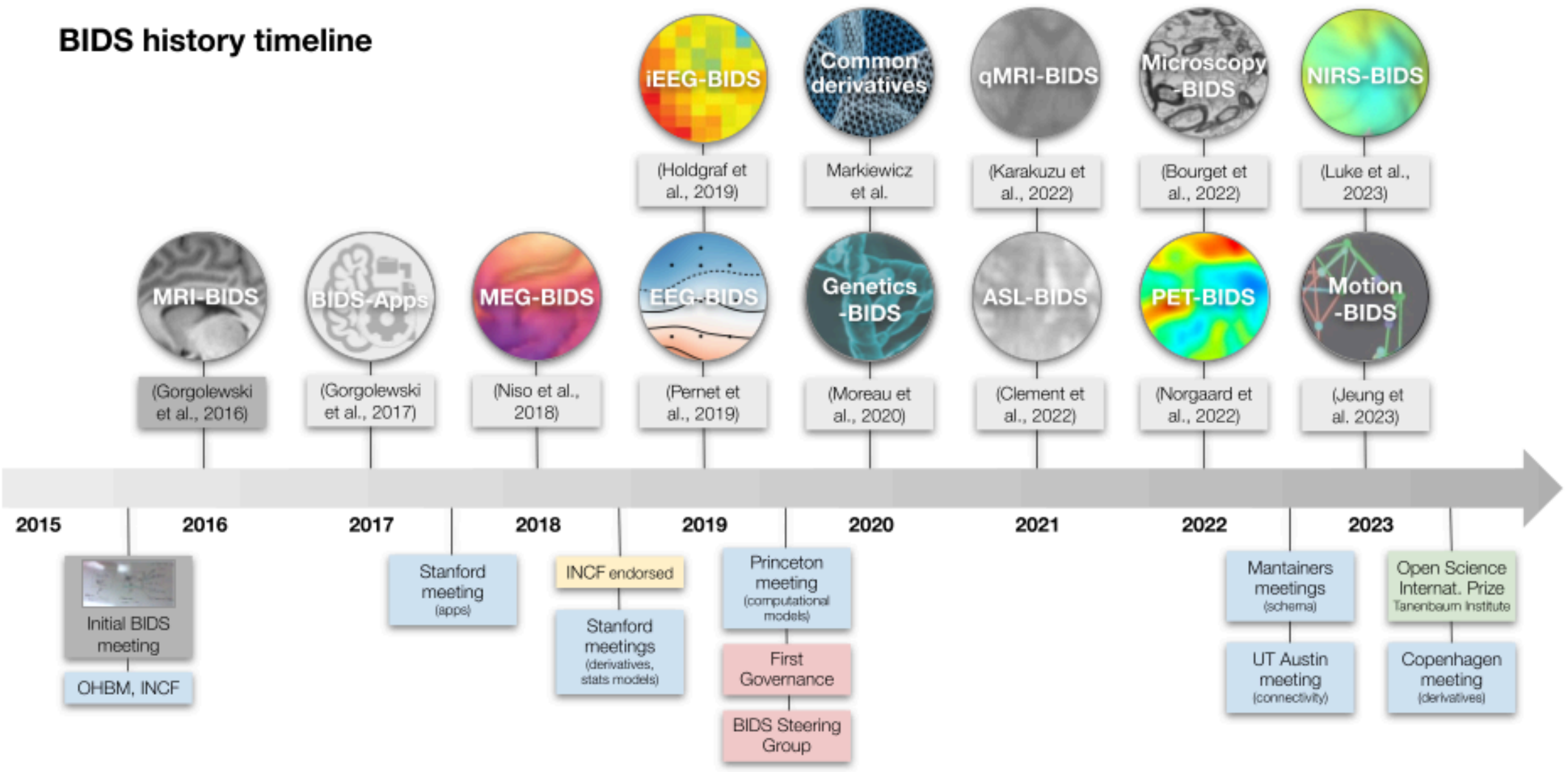

**Figure 2**. A graphical timeline of the historical development of the BIDS project, including important publications, meetings, and other developments.

A timeline of the development of BIDS is presented in Figure 2. The Stanford meeting was held January 27th-30th, 2015, with significant support from the INCF[3]. The in-person attendees were Mathew Abrams, Michel Dumontier, Guillaume Flandin, Chris Gorgolewski, Karl Helmer, David Keator, Camille Maumet, Nolan Nichols, Russ Poldrack, Jean-Baptiste Poline, Ariel Rokem, and Vanessa Sochat; other invitees attending remotely included Satrajit Ghosh, Yaroslav Halchenko, Michael Hanke, David Kennedy, Angie Laird, Tom Nichols, and Jessica Turner. The meeting started with presentations on a number of ongoing relevant projects and their relation to the OpenfMRI use case. The first explicit mention of BIDS came on Day 2, when one of the subgroups was labeled as "Subgroup1: OBIDS (Open Brain Imaging Data Structure) format proposal (derived from the OpenfMRI format)". A photo of a whiteboard drawing (Figure 3) shows the intended separation of standards, with a directory-based format (intended for most users) and a Resource Description Framework (RDF) based format (intended for computationally advanced users); the former is what would become BIDS.

[3] The original agenda and notes from the meeting are available at https://osf.io/kmavh/

Following the meeting in January 2015, there was a substantial effort to develop a set of examples that would be circulated with the initial draft of the specification, based primarily upon datasets from the OpenfMRI database. An additional meeting was held at the OHBM conference in Honolulu in June, 2015 and consequently at the INCF 2015 congress in Cairns, Australia, where both OBIDS and NIDM groups worked together as part of NIDASH to cross-fertilize both standards with basic metadata to describe MRI experiments and data. The BIDS specification draft and 22 example datasets were disseminated to the community for comments on September 21st 2015, along with an early version of the JavaScript-based bids-validator[4], developed by Squishymedia. Squishymedia was a contractor also responsible for primary development of the OpenNeuro archive; it later ended operations in 2021, after which Nell Hardcastle (lead developer on the OpenNeuro project) joined the Stanford team. A NIDASH task force meeting was held in Chicago in October 2015 in advance of the Society for Neuroscience (SFN) Annual Meeting, and a BIDS leaflet[5] was distributed at the SFN meeting to promote the standard to a wider audience.

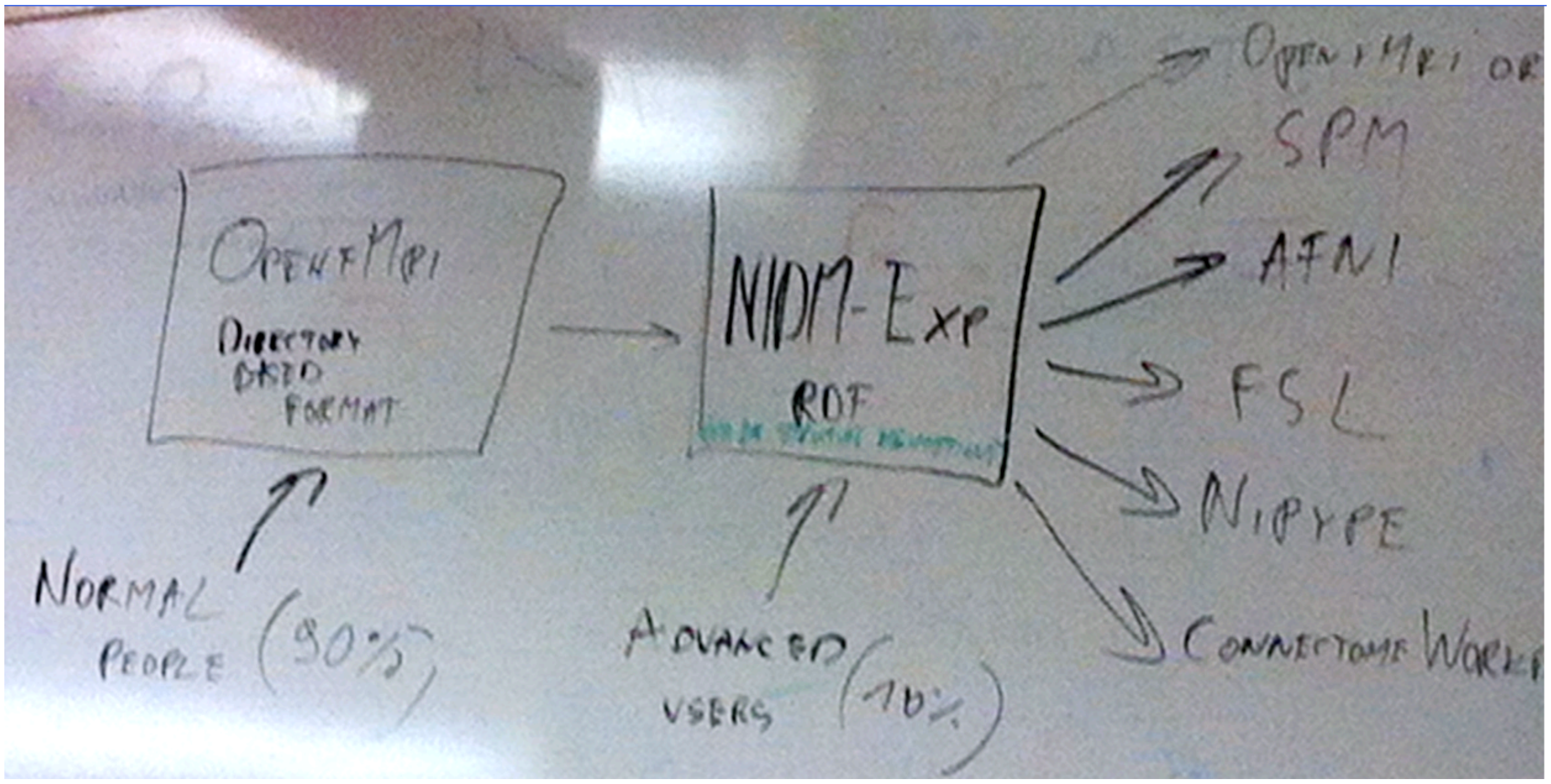

**Figure 3**: A snapshot of the whiteboard at the initial BIDS meeting (January 27-30, 2015), outlining the intended separation of a directory-based format (which would become BIDS) and a formal RDF-based description (which would become NIDM-Experiment).

[4] https://github.com/bids-standard/bids-validator/

[5] A copy available from https://web.archive.org/web/20230519223151/https://neuro.debian.net/_files/brochure_bids.pdf and also listed on https://centerforopenneuroscience.org/engage.

## The establishment of BIDS

The publication of a paper in *Scientific Data* in June 2016 (Gorgolewski et al., 2016) marked the initial official release of the standard. As BIDS developed, a set of principles became enshrined which have been central to subsequent decision-making in the project. The first principle is that *adoption is crucial*. This led to a focus on engaging the relevant research community, and on keeping the standard as close as possible to the practices already in use in the community, thus requiring engineering ingenuity to build a solution from simple existing components while avoiding technical complexity. For example, this led to the focus on human-readable text formats rather than more complicated formats such as RDF or XML (Extensible Markup Language).

The second principle is: *don't reinvent the wheel*. This led to the use of existing standards and file formats whenever possible, including NIfTI, JSON (JavaScript Object Notation), and tab-separated text files (TSV), along with other projects later incorporated, such as the Hierarchical Event Descriptor (HED) annotation system (Robbins et al., 2022). It is important to note the historical trajectory that led to the choice of these particular formats. For example, Digital Imaging and Communications in Medicine (DICOM) has long been the dominant imaging data format within the field of medical imaging. However, due to the overall complexity of DICOM and the early absence of explicit MRI support (with DICOM Working Group 16 on MRI only established in 1998), the academic and research neuroimaging field in the 1990s turned to simpler formats, such as Analyze (Robb et al., 1989). Subsequently, in the early 2000s, this simpler format inspired the creation of the NIfTI format, which quickly became the standard for data exchange and storage in academic neuroimaging research. As advancements were made in MRI scanners, enhancing their capability to export acquired data in DICOM format, tools like dcm2nii and dcm2niix (Li et al., 2016) emerged to convert DICOM series into the simpler and widely supported NIfTI format. However, as NIfTI encapsulates only minimal metadata, compared to the comprehensive metadata represented within DICOM, these tools also needed to export the more complete version of the metadata into separate ("sidecar") files, generally represented using JSON. Currently, over 70 metadata fields in BIDS sidecar files originate from or are based on metadata acquired from DICOM series. DICOM has become the primary format for data export from MRI and PET scanners, which are then converted to NIfTI for interoperability with commonly used software packages.

Finally, the group has focused on the work that will achieve the largest amount of impact, and on not letting the perfect be the enemy of the good; this principle is now referred to informally as the "80/20 rule". This requires admitting that there will always be edge cases that can't be well accommodated by the standard, but that this is fine as long as the standard works well for most people in most circumstances most of the time. Perhaps most important was the community outreach performed by Chris Gorgolewski as the leader of the project, whose dogged pursuit of feedback from a broad group of stakeholders led to a general feeling of inclusiveness and community-mindedness.

One particular idea for incentivizing the use of the standard was to develop software applications that could be easily applied to a BIDS dataset. This idea led to the concept of the *BIDS App* (Gorgolewski et al., 2017), which refers to a containerized software application of a specific analysis pipeline that is aware of the BIDS standard and can thus be applied directly to a BIDS dataset. This model has become highly successful, leading to widely used tools, including fMRIPrep (Esteban, Markiewicz, et al., 2019), QSIPrep (Cieslak et al., 2021)(Torabian et al., 2023), and MRIQC (Esteban et al., 2017), and has almost certainly driven adoption of the standard by researchers who wish to use these software tools[6]. Those early efforts on BIDS Apps also catalyzed the development of interoperability layers for existing tools (e.g., the MNE-BIDS interoperability layer for MNE-Python; (Appelhoff et al., 2019) as well as generic libraries to facilitate interaction with BIDS datasets including PyBIDS (Yarkoni et al., 2019), BIDS-Matlab (Gau et al., 2022), and nilearn (RRID:SCR_001362). In particular, interactions between BIDS and developers of existing tools were nurtured at a series of coding sprints organized by the Stanford Center for Reproducible Neuroscience in 2016 and 2017.

### BIDS Extensions

The initial BIDS standard was focused on specific types of MRI data, but even before its initial release it became clear that there would be a need for extensions to accommodate additional data types. Soon after the establishment of the standard, the BIDS team began to formalize the concept of extending BIDS through “BIDS Extension Proposals” (BEPs), inspired by the Python Enhancement Proposal (PEP) process. The first extension proposals to be intensively discussed and eventually merged into the specification were BEP008 (“Magnetoencephalography”) and BEP007 (“Hierarchical Event Descriptor” HED tags) in BIDS version 1.1, which were quickly followed by BEP006 (“Electroencephalography”) and BEP010 (“intracranial Electroencephalography”) in BIDS version 1.2. As the BEP numbers suggest, there were proposals even before these (e.g., BEPs 001 through 005); however, they progressed more slowly and were merged at later times or are still works in progress; see Table 1 for a list of merged BEPs and https://bids.neuroimaging.io/get_involved.html for full documentation of all BEPs.

The process of developing a BEP was deliberately designed to be as inclusive and with as few technical hurdles for potential contributors as possible. Although the process has become more formal throughout the years and central documentation about the necessary steps and deliverables exists (https://bids-extensions.readthedocs.io), its core has stayed the same: Interested parties may suggest a BEP via the BIDS mailing list or central development repository on Github. After an initial check by a group that has come to be known as the “BIDS maintainers” (see below) of whether the proposed idea broadly fits the scope of BIDS, a BEP receives a number and an entry in the list of ongoing BEPs. The BEP itself is then developed using shared public documents that are advertised via the central BIDS website, and where any person may become a contributor by adding comments and suggestions. Each BEP is headed by a team of "leads" or "moderators," who are responsible for tracking individual contributions, steering discussions, and accepting and rejecting suggestions to the document. The BEP leads are furthermore responsible for organizing virtual discussion rounds (e.g., video calls) with all contributors, and for explicitly reaching out to the broader community working with the relevant methods. Several community members of BIDS that are active to this day have initially found their way into the community by stumbling over a BEP document and offering their feedback and/or minor additions (i.e., simple typo corrections), further highlighting the inclusiveness of the community-driven process.

[6] See https://bids-apps.neuroimaging.io/apps/ for a complete list of currently available BIDS apps.

The first BIDS extension released for a new neuroimaging modality was for magnetoencephalography (MEG) (Niso et al., 2018). The design process was initiated around the creation of the Open MEG Archive (OMEGA; (Niso et al., 2016)) at the Montreal Neurological Institute, for which BIDS was identified as a principled framework to organize the multimodal data. Sylvain Baillet and Guiomar Niso thought that the development and adoption of a data organization standard for MEG would require the active participation of academic software developers in the field. An ad hoc working group first met around the idea of a possible MEG extension to BIDS at the International Conference on Biomagnetism held early October 2016 in Seoul, South Korea[7]. The first elements of specification for MEG-BIDS were shared as a preprint (Niso et al., 2018), which received comments and suggestions from the community. In July 2017, further input from the MEG community was collected through an online poll survey. The poll received 78 international entries, with the most salient results indicating strong interest in a common standard for MEG data organization (~99%) and affirming a willingness to try a MEG-BIDS solution (97%). Although the initial thought was to develop a single BIDS extension encompassing all electrophysiological data modalities, the working group identified that each data type required specific adjustments that would be best addressed by their respective experts. The MEG extension to BIDS introduced substantial innovations that stemmed from the original MRI elements, considering the radically different nature of the instrument technology and time-resolved data types, paving the way for other electrophysiological modalities, such as EEG (Pernet et al., 2019) and iEEG (Holdgraf et al., 2019). With these three electrophysiology extensions (MEG, EEG, iEEG), BIDS became a common structure for a larger user group, including many different types of neuroimaging scientists beyond the original use case of MRI.

| BEP # | Title | Date merged | Publication |
|---|---|---|---|
| BEP001 | Quantitative MRI (qMRI) | 2021-02-23 (v1.5.0) | (Karakuzu et al., 2022) |
| BEP003 | Common Derivatives | 2020-06-11 (v1.4.0) | |

[7] http://www.biomag2016.org/download/program/BIOMAG2016_Poster_Sessions.pdf, Poster Mo-P011

| | | | |
|---|---|---|---|
| BEP005 | Arterial Spin Labeling (ASL) | 2021-02-23 (v1.5.0) | (Clement et al., 2022) |
| BEP006 | Electroencephalography (EEG) | 2019-03-04 (v1.2.0) | (Pernet et al., 2019) |
| BEP007 | Hierarchical Event Descriptor (HED) Tags | 2018-04-19 (v1.1.0) | (Robbins et al., 2022) |
| BEP008 | Magnetoencephalography (MEG) | 2018-04-19 (v1.1.0) | (Niso et al., 2018) |
| BEP009 | Positron Emission Tomography (PET) | 2021-04-22 (v1.6.0) | (Norgaard et al., 2022) |
| BEP010 | intracranial Electroencephalography (iEEG) | 2019-03-04 (v1.2.0) | (Holdgraf et al., 2019) |
| BEP018 | Genetic information | 2020-04-14 (v1.3.0) | (Moreau et al., 2020) |
| BEP030 | Near Infrared Spectroscopy (NIRS) | 2022-10-29 (v1.8.0) | (Luke et al., n.d.) |
| BEP031 | Microscopy | 2022-02-15 (v1.7.0) | (Bourget et al., 2022) |

**Table 1**. Merged BIDS Extensions as of December 2023. A full link of all BEPs (including works in progress) is available at [https://bids.neuroimaging.io/get_involved.html](https://bids.neuroimaging.io/get_involved.html).

## BIDS Derivatives

The initial BIDS standard focused on “raw” data[8]. As a result of the excitement in the community for the BIDS standard, early growth of the project was horizontal, with a number of new data modalities being added over time. However, the need for a standard to organize processed data quickly became apparent, given that the end results of these studies are derived not directly from raw data, but rather through potentially numerous intermediate processing steps, possibly produced by one or more BIDS Apps. Discussions began in 2017 of a “layout file for derivatives", and this led to the further development of the concept of a BIDS Derivatives Standard. In 2018, the US BRAIN Initiative funded a grant to develop BIDS standards for processed data, computational models, and statistical models[9]. This grant funded three meetings held in 2018 and 2019, in which members of the community convened to work on these new standards.

[8] The concept of “raw” data remains controversial, with different researchers having very different interpretations of the term. The initial usage was meant to capture its most common usage in the MRI field, i.e. data as downloaded from the MRI scanner, usually in DICOM format; however, this usage conflicted with the term’s usage in other modalities. At a BIDS Derivatives meeting in Copenhagen, Denmark in 2023, an operational definition of the term “raw” was proposed, such that a “raw BIDS dataset” is one for which there is no source BIDS dataset specified.

The description of processed data is much more challenging than the description of raw data, since there is an almost infinite set of possible processing operations that might be applied to a neuroimaging dataset and outputs that can be generated. Compounding this challenge are differences among practitioners of different modalities (for example, functional and diffusion imaging) for sharing processed results, particularly with regard to what is considered necessary metadata and provenance. At an August 2017 meeting at Stanford, it was agreed to split the BIDS Derivatives specification (BEP003) into a series of BEPs focused on particular modalities or use cases, to allow for independent development on largely independent components. An August 2018 meeting focused specifically on derivatives with a goal of advancing each sub-proposal, establishing common principles, and recombining into a single BEP, leading to a finalized document in December 2018 that was called "BIDS Derivatives Release Candidate 1" that was advertised for implementation. Following the OHBM 2019 meeting, in the absence of Chris Gorgolewski as a primary driver for integration of all derivatives at once (see next section), the sections of BIDS Derivatives were again broken into sub-proposals, with the common principles being the most important target for inclusion into BIDS. These Common Derivatives principles were finally released as part of BIDS 1.4.0 in June 2020.

The initial push for BIDS Derivatives also included two projects aimed at the description of computational artifacts related to neuroimaging data. The first was the BIDS Computational Models framework, which was meant to describe computational models used in neuroimaging data analysis (Poldrack et al., 2019). The Stanford group joined forces with Jonathan Cohen at Princeton University, who was developing a related project called PsyNeuLink[10], and held a meeting at Princeton in April 2019. This was a wide-ranging meeting, which focused on establishing a set of use cases and example implementations that could be used to guide further development. Ultimately, this effort split in two different directions. One effort to develop a generic framework for the specification of a broad set of computational models, led by Jonathan Cohen, Padraig Gleeson, and Tal Yarkoni, became the ModECI Model Description Framework (Gleeson et al., 2023); this is an example where a standard initially developed as part of BIDS became a freestanding separate project. A separate effort led by Petra Ritter aimed at establishing a more constrained framework for the specification of inputs to and outputs from computational network modeling software (such as the Virtual Brain; Sanz Leon et al., 2013; Schirner et al., 2022); this project is currently developing a BEP for review.

Another aim of the BIDS Derivatives project was to develop a framework to describe common statistical models that are applied to neuroimaging data. Statistical modeling of neuroimaging data follows broadly stereotyped patterns, with researchers making a relatively small number of significant decisions within established pipelines. However, the details of the pipelines' construction lead to a high degree of analytical flexibility (Botvinik-Nezer et al., 2020), which is difficult to capture using idiosyncratic user-generated code. The "BIDS Stats Model" specification was created on the observation that a majority of neuroimaging analysis methods can be represented by the General Linear Model (GLM) and common data flow operations. The effort kicked off with an initial document by Chris Gorgolewski, Tal Yarkoni, and Satra Ghosh in September 2016, which became BEP002. Subsequent efforts to develop the specification and the tooling were driven by Neuroscout (de la Vega et al., 2022), which was the first project to use the specification in production. An effort to substantially complete the specification was undertaken at a meeting at Stanford in October 2018, and was subsequently led by Tal Yarkoni, Alejandro de la Vega, and Chris Markiewicz until Yarkoni left academia in 2021. The specification was finalized at a Spring 2022 meeting at the University of Texas and published as a standalone website (https://bids-standard.github.io/stats-models/) with dual goals of creating user-facing documentation, validator, and technical specifications for implementations. Currently, the BIDS Stats Model specification is able to represent a majority of GLM models used in neuroimaging, and is implemented by PyBIDS/FitLins and BIDS-MATLAB/bidspm. Ongoing efforts are being made to develop tools to facilitate user-friendly model specification, as well as integration of the specification with the most popular neuroimaging analysis packages.

---

[9] Original grant proposal and meeting summaries at https://osf.io/c3dgx/
[10] https://princetonuniversity.github.io/PsyNeuLink/

Other efforts have worked to add specifications for specific forms of derivative data across multiple modalities. One example is the BIDS Connectivity project, a BRAIN Initiative project led by Franco Pestilli of the University of Texas that aims to extend the BIDS standard to describe research objects commonly used in experiments related to brain connectivity. The BIDS connectivity extension(s) project comprises most major neuroimaging data modalities (sMRI, fMRI, DWI, PET, EEG, iEEG, and MEG) and covers data products pertaining to a broad range of analyses, including structural and functional connectivity matrices, seed-based connectivity maps, networks based on dimensionality reduction and tractograms and tractometry. The BIDS connectivity project is the first example of a large-scale coordination effort spanning multiple BEPs, with at least five BEPs encompassing different derivative types being advanced in a coordinated manner.

### Weathering an existential crisis

BIDS was envisioned from the beginning as a community project, but in practice its leadership fell heavily on the shoulders of its founder, Chris Gorgolewski. Chris's strong leadership role in the BIDS community ended with his departure from academia in March 2019, and the resulting leadership vacuum led to an existential crisis within the BIDS community, which had by then grown substantially. This discussion came to a head at a BIDS Town Hall meeting organized at the OHBM 2019 meeting in Rome[11]. Following an overview of the BIDS project, there was a spirited discussion about the future leadership of BIDS. This discussion continued on the BIDS Standard GitHub repository[12]; a particularly impactful argument was made by Kirstie Whitaker in favor of community governance and against the "tyranny of structurelessness" (Freeman, 1972) in which a small number of voices come to dominate the discussion.

---

[11] Slides available at https://osf.io/kmavh/
[12] https://github.com/bids-standard/bids-specification/pull/104

In the wake of the OHBM meeting, Tom Nichols proposed the idea of a BIDS Steering Group, which would be elected based on a vote by the BIDS community. In parallel, a plan for community governance was developed by members of the community (see [https://bids.neuroimaging.io/governance](https://bids.neuroimaging.io/governance)). The development of this process was led substantially by Franklin Feingold from the Stanford team, who served as the principal project manager for BIDS from 2018 through 2022. The first election was held in October 2019. The winning slate was chaired by Guiomar Niso including Melanie Ganz, Robert Oostenveld, Russ Poldrack, and Kirstie Whitaker, a team that represented the diversity of neuroimaging modalities already then in BIDS, including MRI, PET, MEG, and EEG. In 2021 an election was held to replace Kirstie Whitaker, which resulted in the election of Ariel Rokem. In 2022, an election was held to replace Melanie Ganz and Russ Poldrack, which resulted in the election of Yaroslav Halchenko and Cyril Pernet. In 2023, an election was held to replace Guiomar Niso and Robert Oostenveld, which resulted in the election of Dora Hermes and Camille Maumet. The continued successful operation of the community in the absence of its founder demonstrates the strength of the community governance model.

The governance process also formalized the role of the BIDS maintainers. This was spurred by the move of the primary BIDS specification document from a collaboratively edited document to a website generated from a version-controlled GitHub repository, which increased the technical barriers to contribution. The initial maintainers' group consisted of Stefan Appelhoff, Franklin Feingold, Ross Blair, and Chris Markiewicz, who took responsibility for managing the repository and validator and facilitating contributions from users less familiar with working in GitHub. The work of the maintainers group includes social infrastructure, such as the BIDS website and social media, as well as running the steering group elections. The maintainers group remains a self-selected set of contributors that collectively make infrastructure decisions around the specification, validator, and example datasets, facilitate additions to the standard, and advise the steering group. Taylor Salo and Remi Gau joined the group in 2020, Anthony Galassi and Eric Earl in 2021, and Christine Rogers, Nell Hardcastle and Kim Ray in 2023.

One key strength of BIDS is the engineering acumen of the scientists involved, which is reflected in the way that the standard is now rendered for public consumption. Initially the published standard for the web site and a machine-readable JSON schema used by the JavaScript validator were kept in sync manually. The development of a generic, declarative schema to describe the standard (described further below) allowed the BIDS maintainers and other contributors to develop tools that generate some crucial specification text and tables directly from the very same schema as the validator and therefore avoid conflicts between the published standard and the validator. The unified schema also allowed downstream tools, such as HeuDiConv (RRID:SCR_017427), to avoid hardcoding the standard, thus making them also more robust to changes to the BIDS standard.

## The present state of BIDS

At present, BIDS is a highly successful example of a community-driven standard for data organization; to our knowledge, it is one of the only grass-roots standards to have gained such broad acceptance within its field. This community support was recognized by the INCF when they endorsed BIDS as a best practice standard in 2018, and re-endorsed it in 2021. Not only has the adoption of BIDS continued to grow, but BIDS itself has grown to reflect developments in the field.

A massive amount of data is now shared in the BIDS format. The OpenNeuro archive (as of June 2023) shares data for more than 34,000 individuals from more than 850 BIDS datasets; Figure 3 shows the consistent increase in the size of this database over time. The ABCD-BIDS Community Collection (Feczko et al., 2021) shares a BIDS version of the ABCD dataset that includes longitudinal data from 11,877 children.

Another lens into BIDS usage comes from the statistics collected by the MRIQC BIDS app, which stores telemetry information about each MRI run (unless the user opts out) that is then shared via an open web API (Esteban, Blair, et al., 2019). Figure 4 shows the cumulative number of unique images (distinguished based on a checksum of each image) submitted to the MRIQC web API between 2018 and June 2023, which demonstrates a sustained and consistent growth in the number of datasets converted to BIDS. Given that these datasets represent only a subset of the total number of BIDS datasets, we would estimate that there are likely to be well over 100,000 datasets with millions of images that have been converted to BIDS at this point. These results highlight the fact that BIDS is being used by a significant number of researchers in the community.

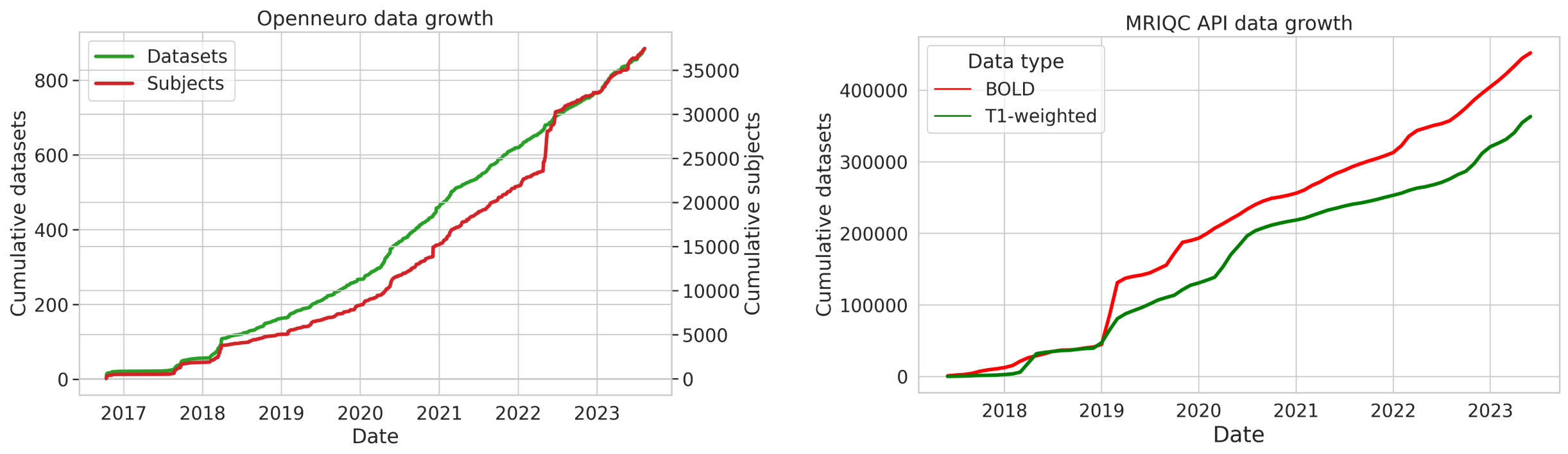

**Figure 4**. Growing usage of BIDS over time. Left: Growth of the OpenNeuro database since its inception in 2017, adapted from (Markiewicz et al., 2021). Right: Cumulative number of unique T1-weighted anatomical (t1w) and BOLD images from BIDS datasets submitted to the MRIQC web API (Esteban, Blair, et al., 2019) from 2018 to June 2023. Source data and code to generate figures available at [https://osf.io/x7fh8/](https://osf.io/x7fh8/).

BIDS has also enabled a number of important data integrations that grow community adoption via the software ecosystem. For example, the cloud platforms brainlife.io (Hayashi et al., 2023) and nemar.org (Delorme et al., 2022) have utilized BIDS to provide ready-to-use data services.

The BIDS standard provides a common data format for ingestion and exchange; combined with DataLad (Halchenko et al., 2021) as a transfer mechanism, OpenNeuro datasets are made seamlessly available to brainlife.io and NEMAR users. Further, the BIDS Apps specification (Gorgolewski et al., 2017) enabled BIDS-aware platforms such as brainlife.io to make third-party BIDS Apps available to a large community of users. Other open-source platforms that integrate BIDS support, such as LORIS (Das et al., 2011) and CBRAIN (Sherif et al., 2014), also contribute open tools and workflows to onboard new user groups to BIDS (Rogers et al., 2022) as a basis for research collaboration. To take one final example, another BRAIN Initiative® data archive, DANDI (RRID:SCR_017571), has effectively employed BIDS in harmony with standards from other subfields, such as Neurodata Without Borders (Teeters et al., 2015) and OME-Zarr (Moore et al., 2023), to facilitate integration across diverse domains of neuroscience data.

One major achievement of the BIDS community in recent years has been the development of a machine-readable schema to represent the standard. The early releases of BIDS consisted of a specification written in English and a validator written in JavaScript, which aspired to have a validation procedure for each English rule. As the standard grew to include more data types, the number of people with the expertise needed to adapt and maintain these parallel representations became vanishingly small. At the same time, each tool implementing BIDS recommendations resulted in yet another representation of the standard that required updating. These difficulties led to an effort to create a declarative (i.e., non-procedural) schema that is maintained as part of the specification document (Figure 5). This has four important benefits. First, it allows much easier inclusion of new elements to the standard (such as new BEPs). Second, it enables the consistent implementation of the validator across multiple languages. Third, it makes it possible for the validator and any downstream tool using the schema to provide handling of the BIDS dataset specific to its BIDS version. Fourth, the inclusion of the schema with the specification encourages contributors who wish to propose a new rule to consider the difficulty of expressing that new rule within the schema. The schema-based validator has been implemented and is currently being tested, and is expected to replace the original validator in late 2023. It has already been adopted by the DANDI and OpenNeuro data archives, both of which contributed to the efforts to develop the schema and schema-based validator.

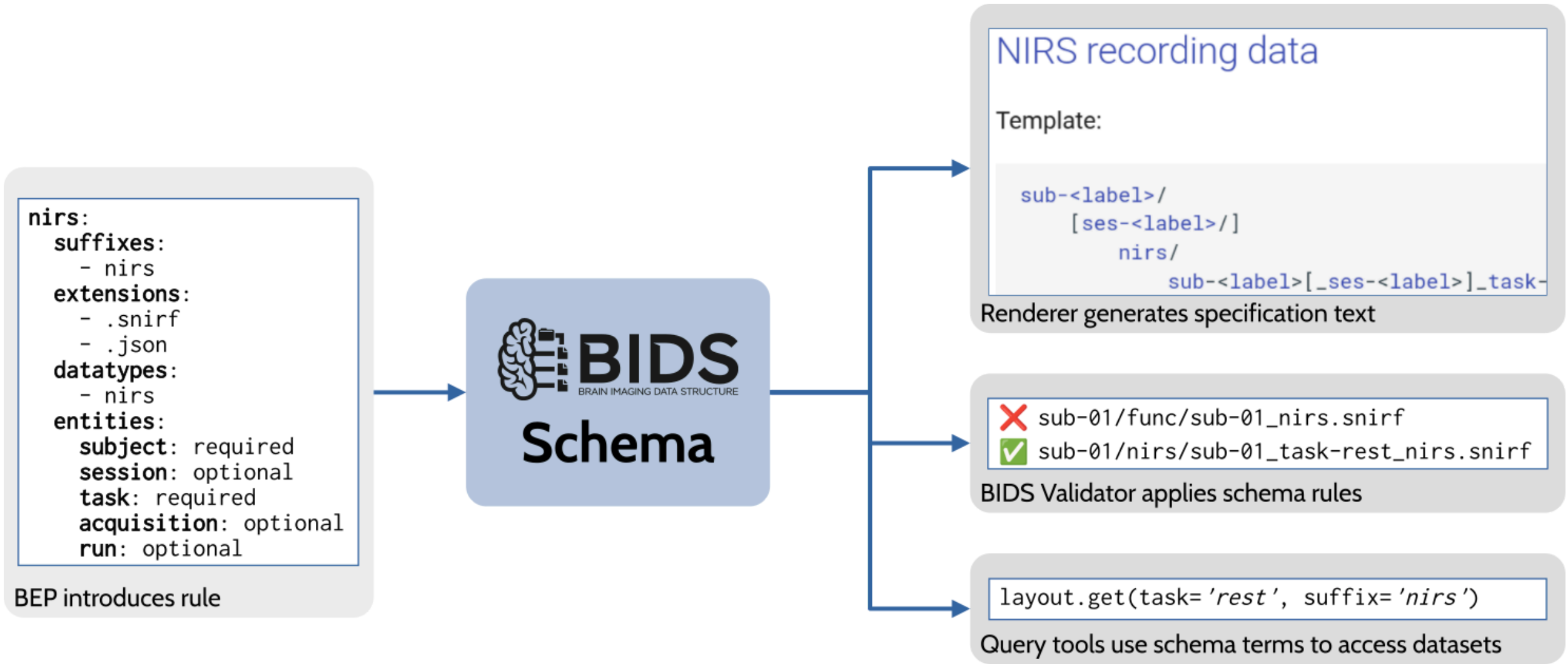

**Figure 5.** Overview of BIDS Schema usage. In this example, BEP 030 (NIRS) introduces a file naming rule to the schema as part of the BEP process. The schema rule is used to render a file naming template in the specification. The BIDS Validator uses the rule to identify valid NIRS data files while rejecting improperly named files. Finally, third-party tools, such as a query library, may ingest the updated schema to automatically gain access to new features of BIDS.

## The future of BIDS

As BIDS nears its tenth anniversary, its success has also led to increasing recognition of the limitations of the existing framework. This has in turn driven a growing discussion regarding the need for a new major version of BIDS ("BIDS 2.0") that would introduce changes that are incompatible with the existing BIDS framework. A dedicated repository (https://github.com/bids-standard/bids-2-devel/issues) is collecting issues, a subset of which will be chosen for BIDS 2.0. It is likely that the discussion of a new version will continue over the next few years, given the significant difficulty that breaking changes would impose on tool developers.

As an example, one contentious issue that would require resolution in any successor specification is the "inheritance principle". This allows metadata to be specified at a higher level in the data hierarchy, and be "inherited" by multiple data files to which it is applicable. On one hand, this introduces a degree of complexity to user comprehension and software interaction with the BIDS specification and BIDS datasets; on the other, it collapses redundant information in a manner that is faithful to the hierarchical nature of the metadata at hand, which becomes increasingly pertinent for datasets of increasing complexity. Many such situations exist where identification of an issue or contention would not have been possible without hands-on experience within the BIDS ecosystem.

Another major challenge for the BIDS community is to decide where the standard should end. BIDS has already extended beyond what is classically considered as "neuroimaging data", through the Microscopy extension (BEP031) and other extensions including genetic information (BEP018), eye tracking (BEP020), and movement data (BEP029). The success of BIDS has garnered interest from many researchers in developing related standards under the BIDS umbrella, but continued expansion of the scope of the standard also threatens to increase its complexity to a degree that it becomes difficult to change.

The question of where the standard ends is particularly relevant to continued standardization of BIDS Derivatives and their interaction with software tools.

1. *Hierarchical complexity*: The filesystem structure of the BIDS standard was designed to meet the requirements of BIDS Raw data, with data files arranged in an immutable hierarchy across datasets, then subject, (optionally) session, and finally imaging modality. While there may be tremendous flexibility in file naming within this structure, the hierarchy itself is inflexible. This did not pose an issue for BEP003 Common Derivatives, as each proposed derivative was a standalone piece of data derived from raw data from a single modality. However there are future prospects where this will no longer be the case, such as derivatives that are the result of explicitly multi-modal analysis, or data hierarchies more complex than that aforementioned (such as the result of a model fit that is spread across multiple data files), which cannot be faithfully represented in a BIDS structure with inheritance under the current standard.

2. *Derivatives as inputs*: From their inception, BIDS Apps were designed on the premise of taking as input a BIDS Raw dataset, and producing as output a final set of computed derivative data. The interest in BIDS Derivatives has been principally in facilitation of the sharing and unambiguous interpretation of such data. In parallel, a number of BIDS Apps have implemented the ability to identify and utilize the derivatives already computed by some other application rather than duplicating those requisite calculations internally. Combining these concepts presents an opportunity for the construction of large, complex, multi-modal processing pipelines incorporating disparate softwares. Just as many existing BIDS Apps are based on constructing and executing a directed acyclic graph of underlying individual commands from different neuroimaging software packages (Gorgolewski et al., 2017), larger analysis pipelines could be based on constructing and executing a directed acyclic graph of underlying BIDS Apps, with BIDS Derivatives serving as the means of translation between those Apps.

3. *Existing toolchains*: A large proportion of neuroimaging analyses are performed using one of a small set of existing software packages. Many of these have long-standing file layout schemes and tools designed to operate on their own specific sets of derivatives. Conformance of such packages to a completely new data structure may therefore be too high an expectation. There are two key ways in which this problem should be considered. Firstly, any derivatives extension proposal should ideally have accompanying it a software tool for performing bidirectional conversions between the outputs of one or more major toolchains and the proposed specification; this would facilitate generation of conforming data by users and BIDS App developers alike, and subsequent manipulation / visualization of shared data using the originating toolchain.

Secondly, an alternative approach would be to describe in the specification a way to encapsulate those results in the native toolchain format and merely annotate them with agreed common terms, an approach with precedents in BEPs 015 (Derivative mapping files) and 035 (Mega-analysis using non-compliant derivatives), and which may dovetail with BEP 028 (Provenance). This would permit *post hoc* annotation of derivatives that predate BIDS Derivatives and/or are generated by tools outside of the BIDS ecosystem; and while it may sacrifice ultimate filesystem convention conformity, it would not preclude the ongoing pursuit of such.

Another direction for future efforts is further integration with other related standards. Whereas the NIfTI file format has become standard within the MRI research community, DICOM has grown into the industry standard for a wide range of imaging modalities (such as physiology, etc.), while addressing the many shortcomings that had originally turned the neuroimaging research community towards simpler formats. As a result, many standardization efforts have been duplicated. As DICOM is the industry standard and more data will be arriving in DICOMs, coordination with developments in DICOM could help to ensure more rapid adoption of new imaging sequences and even modalities into the BIDS standard, which provides an umbrella organization at the study level. Continued work with related standards such as NWB and OME-Zarr will also be important to avoid unnecessary duplication of effort or conflicting recommendations for datasets and archives on the interface of neuroimaging and electrophysiology/microscopy.

The funding of continued BIDS development and maintenance remains a challenge. The BRAIN Initiative has funded many of the developments of BIDS, either directly (for BIDS development projects) or indirectly (through funding to data archives that have relied upon and contributed to BIDS). Much of the work to develop and maintain the standard is performed by the BIDS Maintainers; many of the maintainers are currently supported by related grants, which endangers the project given that grants usually have time windows of three to five years. The establishment of a foundation to support BIDS could be a useful future development.

## Lessons Learned

Given the demonstrable success of BIDS, it is useful to ask: What lessons can be learned that might be useful for other standards projects? The first important point to acknowledge is that the success of any particular project derives in part from pure luck, so one should not overfit too heavily to the following. Nonetheless, we believe that there are several potentially important lessons to be learned.

### Main factors in the success of BIDS

**Absence of existing solutions.** Many standardization projects commence on the premise that there are multiple existing competing standards in that domain, each with their own strengths and weaknesses, and a new standard is sought that inherits more desirable attributes, takes into account lessons learned from those prior standards, and achieves more widespread adoption. The BIDS project was quite unique in that, within the neuroimaging community, there was effectively universal acceptance of ad hoc organization of neuroimaging data beyond the formatting of individual files. The benefits that were to be inherited through the prospect of field-wide harmonization were therefore implicitly attributed to BIDS itself.

**Clear use cases.** One of the main initial motivations to develop BIDS was to allow ingestion of datasets into data repositories at scale without the need for human curation. This not only provided a practical problem that defined the solution (as reflected by the project principles: “adoption is crucial”, “don’t reinvent the wheel”, and “80/20 rule”), but also enabled the team to fund the initial development of the standard via the OpenNeuro project. Subsequently, BIDS apps provided use cases for adoption by end-users, by enabling them to effortlessly process BIDS datasets using high-quality software tools.

**Solving a common end user problem.** Standardizing data organization was not only beneficial for data repositories and software platforms, but also for individual labs and projects. Even if data are never shared publicly, having a standard data organization scheme helps researchers within a lab group work together, and enables future reuse of the data within the lab. The development of BIDS absolved individual groups of the need to develop, document, and implement their own individual schemes, and provided tools (such as the validator and conversion tools) to help with this. Moving the project away from purely “open science” and “data sharing” framing by dropping “Open” from the name was crucial to reinforce the message that BIDS was not only for external sharing.

**Low technical barrier to entry**. The practices adopted by BIDS do not require sophisticated tools beyond those already widely in use in most laboratories, enabling rapid and widespread adoption. At the same time, community-based efforts arose to provide introductory materials to explain the core concepts of BIDS to a broad audience, such as the BIDS Starter Kit[13]. It is also worth acknowledging that community forums such as NeuroStars[14] have played a role in educating users as well as sharing best practices that fall outside the scope of the standard itself.

**Maturity and the size of the field.** When the BIDS efforts started, human neuroimaging was more than 30 years old. Standard patterns across experimental design and data types had already emerged, and common file formats were already widely adopted in some subfields (such as NIfTI within the MRI community). At the same time the overall size of the community (several thousand scientists, several hundred labs) allowed the team to gather detailed feedback from a significant portion of the community and required small overall alignment effort.

**Broad financial and institutional support.** BIDS has the benefit of early support by INCF (which supported a set of meetings of the neuroimaging data sharing (NIDASH) task force that included the first meeting at Stanford), and major funding from the Laura and John Arnold Foundation subsequently provided support for the early development of BIDS. Since then, BIDS development has been supported by grants from a number of institutions (NIMH, NSF, Novo

[13] https://bids-standard.github.io/bids-starter-kit/
[14] https://neurostars.org/

Nordisk Foundation, French National Research Agency) to a number of different investigators[15], which has broadened the base for support of the project and ensured that it didn't rely too heavily on any one particular grant or researcher.

**In person meetings with a diverse set of participants.** Key aspects of BIDS were drafted over many in-person meetings with participants traveling long and far to attend, and various BEPs have been started and developed spontaneously at conferences, workshops and hackathons. These in person meetings were key to brainstorming the different aspects of the standard and chart the work that happened later asynchronously.

**Open doors without "death by consensus".** BIDS managed to finely balance having an open structure and listening to feedback from many members of the neuroimaging community, but at the same time avoiding trying to please everyone and creating a standard that was too flexible to be usable. An example of this was denying early calls to allow the MINC file format in addition to NIfTI; despite the technical superiority of MINC, this would have made supporting the standard by any software tool much harder and diminished its adoption. Achieving this outcome required deft leadership by Chris Gorgolewski, which relied heavily upon his trusted position in the community.

### Stumbling points for the BIDS project

A number of challenges have arisen during the development of BIDS, which developers of other standards can also possibly learn from.

**Delayed adoption by large databanks.** With the exception of OpenNeuro, BIDS was not adopted by any large databases or consortia during its early development. Gaining the support of a large prospective data sharing project (such as UK Biobank or Human Connectome Project) at launch would have made it stronger and helped promote it. Instead, researchers have created "BIDS-ified" versions of these datasets, such as the ABCD-BIDS Community Collection (Feczko et al., 2021). A major challenge with a community-driven project like BIDS is the relatively slow pace of development (due to the need for community input and convergence), which can conflict with the desire of large projects to solve their problems quickly without constraints from outside. It would be useful in the future for funding agencies to require large data-generation projects to employ community standards in order to enhance the FAIRness of the resulting datasets.

**Challenges of BIDS conversion.** Conversion of new datasets into the BIDS format has been, and often remains, a challenge for many researchers, which has limited even wider adoption of the standard. Converters have proliferated[16], covering more use cases but also potentially increasing the burden on new users to choose which conversion tool to adopt. Earlier concerted efforts on developing more easily usable conversion tools could have increased early adoption.

---

[15] Full list of funding is available at https://bids.neuroimaging.io/acknowledgments.html
[16] A current list of BIDS converters is available at https://bids.neuroimaging.io/benefits#converters

**Lack of a machine-readable standard.** For many years, there was no common machine-readable instantiation of the standard. This led to misalignment between the standard, the validator, and the documentation, and made the implementation of changes in the standard difficult and time-consuming. This has been addressed by the schematization of the standard, but earlier development of a machine-readable standard may have improved the development workflow.

**Challenges of BEP management**. While BIDS Extension Proposals (BEPs) will remain the main driver of future BIDS development, they bring unique challenges for community management. Some BEPs may see development stagnate when faced with low developer availability or as project requirements are clarified. In the best case, this relative dormancy marks a clear growth point for the project, as developers realize that more work is necessary to clarify the BEP scope. In the worst case, BEPs may be abandoned by the proposing team; however, if this is telegraphed appropriately, other community members can choose to step in. Other more conceptual challenges include ensuring the same level of community consultation across BEPs. BEP leads who have not previously participated in BIDS development may not be familiar with its governance process. This is particularly concerning if BEPs are preemptively yoked to traditional incentives, with BEP leads committing to firm timelines for associated publications, grant deliverables, or graduate degree progress. Ensuring that BEPs maintain the same level of community consultation while sustaining the engagement of domain researchers motivated to develop BEPs is an ongoing challenge for BEP management.

**Geographical diversity and inclusivity**. Figure 6 shows the institutional locations of all authors on the present paper. This map highlights the fact that BIDS has received contributions from a diverse group of locations across the United States, Western Europe, and Australia. At the same time, there is a notable lack of contributions from researchers in other parts of the world. A goal for the future development of BIDS is to include researchers from these parts of the world that are not currently well-represented in the community.

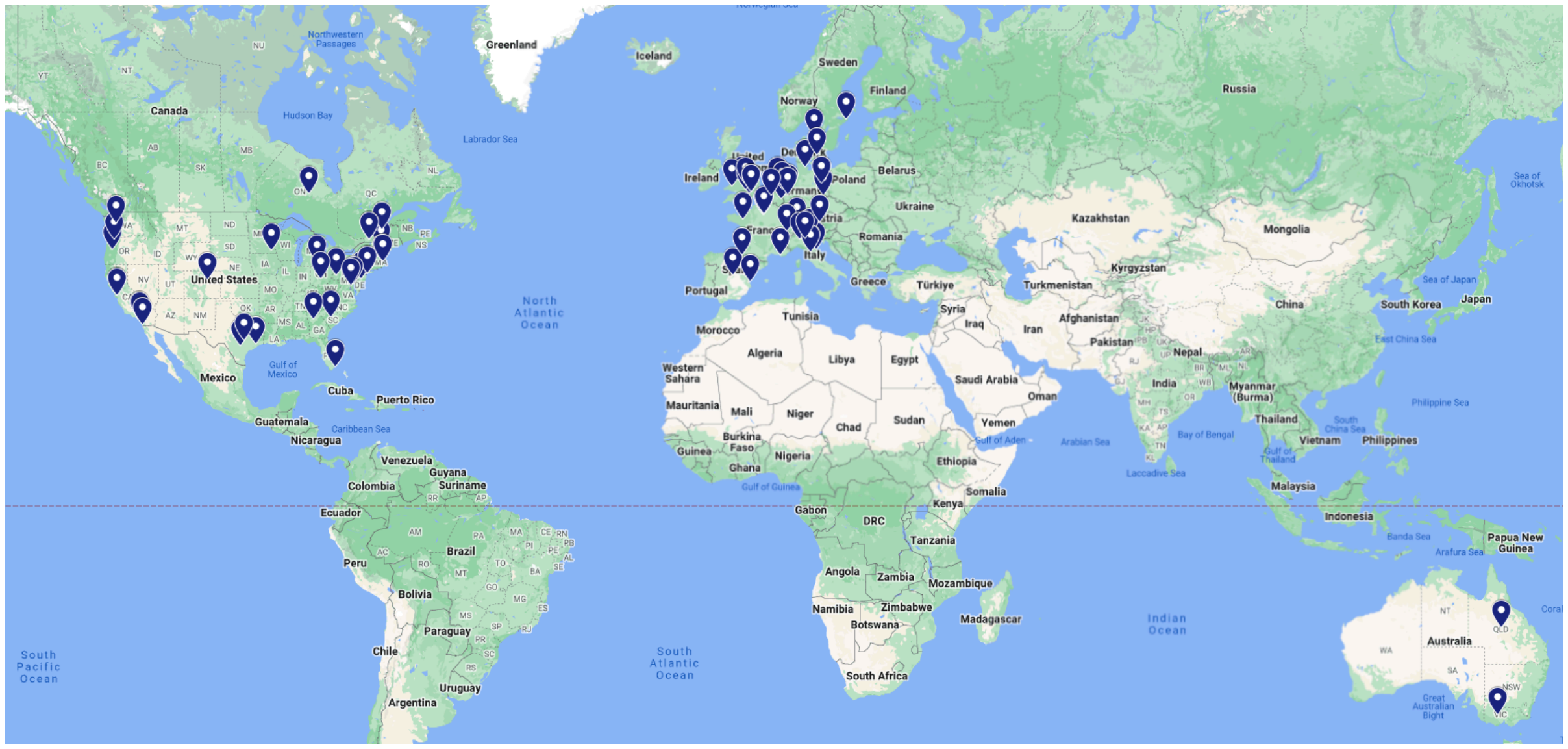

**Figure 6**. A world map of the institutional locations of all coauthors on the present paper.

## Conclusions

Few of us would have envisioned in 2015 that BIDS would be as successful as it has been, or that it would weather the departure of its initial founder so robustly. This success is a testament to the sustained efforts of the large number of individuals who have contributed in many different ways to the community and surrounding ecosystem of tools and data. We hope that BIDS can serve as a demonstration that communities of researchers can effectively develop standards that are essential to enable the effective and FAIR sharing of research data.

## Acknowledgments

Development of the BIDS Standard has been supported by the International Neuroinformatics Coordinating Facility, Laura and John Arnold Foundation, National Institutes of Health (R24MH114705, R24MH117179, R01MH126699, R24MH117295, P41EB019936, ZIAMH002977, R01MH109682, RF1MH126700, R01EB020740), National Science Foundation (OAC-1760950, BCS-1734853, CRCNS-1429999, CRCNS-1912266), Novo Nordisk Fonden (NNF20OC0063277), French National Research Agency (ANR-19-DATA-0023, ANR 19-DATA-0021), Digital Europe TEF-Health (101100700), EU H2020 Virtual Brain Cloud (826421), Human Brain Project (SGA2 785907, SGA3 945539), European Research Council (Consolidator 683049), German Research Foundation (SFB 1436/425899996), SFB 1315/327654276, SFB 936/178316478, SFB-TRR 295/424778381), SPP Computational Connectomics (RI 2073/6-1, RI 2073/10-2, RI 2073/9-1), European Innovation Council PHRASE Horizon (101058240), Berlin Institute of Health & Foundation Charité, Johanna Quandt Excellence Initiative, ERAPerMed Pattern-Cog, and the Virtual Research Environment at the Charité Berlin – a node of EBRAINS Health Data Cloud.

### Declaration of Competing Interests

GR is an employee of Invicro. No other authors have competing interests to declare.

### Use of generative AI language tools

Generative AI tools were not used in the generation of the manuscript text, but were used in the formatting of the author affiliations and generation of code for the figures.

### Data and Code Availability

All data and code used to generate the figures in this paper are available at https://osf.io/x7fh8/ under a permissive open source license.

### Contributor Roles Taxonomy (CRediT) statement

Authorship was offered to any individual who has contributed to the development of BIDS. Authors were asked to denote their contributions to the project according to the Contributor Roles Taxonomy (CRediT). Role abbreviations. C: Conceptualization, DC: Data curation for initial use and later re-use.), FA: Funding acquisition, M: Methodology, PA: Project administration, R: Resources, S: Software, SU: Supervision, V: Validation, W: Writing – original draft, RE: Writing – review & editing.

Russell A. Poldrack: C,FA,M,PA,R,SU,W. Christopher J. Markiewicz: C,M,PA,S,SU,V,W. Stefan Appelhoff: C,DC,M,PA,R,S,SU,V,W. Yoni K. Ashar: S,V,RE. Tibor Auer: C,DC,M,S,V,RE. Sylvain Baillet: C,DC,FA,PA,R,S,SU,RE. Shashank Bansal: S,V,RE. Leandro Beltrachini: M,RE. Christian G. Bénar: SU,RE. Giacomo Bertazzoli: C,M,RE. Suyash Bhogawar: C,DC,M,R,S,SU,V,RE. Ross W. Blair: C,PA,S,RE. Marta Bortoletto: C,RE. Mathieu Boudreau: C,M,SU,RE. Teon L. Brooks: M,S,RE. Vince D. Calhoun: FA,M,S,RE. Filippo Maria Castelli: DC,S,RE. Patricia Clement: M,PA,R,SU,RE. Alexander L. Cohen: S,V,RE. Julien Cohen-Adad: M,S,SU,RE. Sasha D'Ambrosio: DC,R,S,RE. Gilles De Hollander: C,DC,PA,R,SU,RE. Alejandro De La Vega: PA,S,SU,V,W. Arnaud Delorme: C,DC,M,R,S,W. Orrin Devinsky: FA,R,RE. Dejan Draschkow: PA,R,SU,RE. Eugene Paul Duff: C,M,RE. Elizabeth Dupre: PA,S,RE. Eric Earl: C,DC,M,PA,S,RE. Oscar Esteban: C,DC,S,V,RE. Franklin W. Feingold: PA,RE. Guillaume Flandin: C,M,R,S,RE. Anthony Galassi: DC,S,SU,RE. Giuseppe Gallitto: C,M,RE. Melanie Ganz: C,FA,PA,R,SU,RE. James Gholam: C,M,RE. Satrajit S. Ghosh: C,DC,FA,R,S,SU,RE. Alessio Giacomel: C,DC,S,RE. Ashley G. Gillman: S,RE. Padraig Gleeson: V,RE. Alexandre Gramfort: FA,R,S,V,RE. Samuel Guay: R,S,RE. Giacomo Guidali: M,RE. Yaroslav O. Halchenko: C,FA,R,S,RE. Daniel A. Handwerker: M,RE. Nell Hardcastle: DC,S,W. Peer Herholz: DC,S,RE. Dora Hermes: DC,M,R,RE. Christopher J. Honey: R,RE. Robert B. Innis: DC,FA,PA,R,RE. Horea-Ioan Ioanas: R,S,V,RE. Andrew Jahn: V,RE. Agah Karakuzu: C,M,PA,R,RE. David B. Keator: S,RE. Gregory Kiar: C,M,S,RE. Balint Kincses: C,RE. Angela R. Laird: DC,R,W. Jonathan C. Lau: C,V,RE. Alberto Lazari: DC,RE. Jon Haitz Legarreta: S,RE. Adam Li: DC,S,V,RE. Xiangrui Li: S,RE. Bradley C. Love: M,RE. Hanzhang Lu: M,R,S,RE. Camille Maumet: C,FA,M,SU,RE. Giacomo Mazzamuto: C,DC,S,RE. Steven L. Meisler: S,RE.

Mark Mikkelsen: C,M,RE. Henk Mutsaerts: C,M,PA,SU,RE. Thomas E. Nichols: C,FA,M,R,SU,RE. Aki Nikolaidis: R,RE. Gustav Nilsonne: PA,RE. Guiomar Niso: C,DC,M,PA,R,S,SU,W. Martin Norgaard: C,DC,M,PA,R,S,SU,RE. Thomas W. Okell: M,R,S,RE. Robert Oostenveld: C,DC,M,PA,R,S,SU,RE. Eduard Ort: S,RE. Patrick J. Park: C,R,S,RE. Mateusz Pawlik: S,RE. Cyril R. Pernet: C,DC,M,PA,S,SU,W. Franco Pestilli: C,FA,M,PA,R,SU,RE. Jan Petr: C,DC,S,SU,V,RE. Christophe Phillips: C,PA,SU,RE. Jean-Baptiste Poline: C,M,R,SU,W. Luca Pollonini: PA,SU,RE. Pradeep Reddy Raamana: C,DC,S,RE. Petra Ritter: C,DC,FA,M,SU,V,RE. Gaia Rizzo: C,M,RE. Kay A. Robbins: C,DC,M,R,S,V,RE. Alexander P. Rockhill: R,S,V,RE. Christine Rogers: DC,S,RE. Ariel Rokem: C,DC,FA,PA,R,S,SU,RE. Chris Rorden: DC,M,S,V,RE. Alexandre Routier: C,RE. Jose Manuel Saborit-Torres: C,DC,FA,S,RE. Taylor Salo: R,S,W. Michael Schirner: C,DC,M,RE. Robert E. Smith: S,W. Tamas Spisak: C,M,RE. Julia Sprenger: C,S,RE. Nicole C. Swann: FA,SU,RE. Martin Szinte: R,S,RE. Sylvain Takerkart: C,FA,S,SU,RE. Bertrand Thirion: C,S,RE. Adam G. Thomas: FA,PA,R,S,SU,RE. Sajjad Torabian: C,M,S,V,RE. Gael Varoquaux: C,S,RE. Maria De La Iglesia Vaya: C,DC,FA,R,S,RE. Bradley Voytek: S,RE. Julius Welzel: C,R,RE. Martin Wilson: DC,S,SU,RE. Krzysztof J. Gorgolewski: C,DC,FA,M,PA,R,S,SU,V,W.